\documentclass[aps,pre,amsmath,amssymb,reprint,onecolumn,floatfix,longbibliography,notitlepage]{revtex4-1}

\usepackage{bm}
\usepackage[dvips]{graphicx}

\begin{document}

\title{
Hysteretic behavior between quasi-two-dimensional flow and three-dimensional flow\\
 in forced rotating turbulence
}

\author{Naoto Yokoyama}
\email{yokoyama@kuaero.kyoto-u.ac.jp}
\affiliation{Department of Aeronautics and Astronautics, Kyoto University, Kyoto 615-8540, Japan}

\author{Masanori Takaoka}
\email{mtakaoka@mail.doshisha.ac.jp}
\affiliation{Department of Mechanical Engineering, Doshisha University, Kyotanabe 610-0394, Japan}

\date{\today}

\begin{abstract}
 Conflict between formation of a cyclonic vortex and isotropization
 in forced homogeneous rotating turbulence
 is numerically investigated.
 It is well known that
 a large rotation rate of the system induces columnar vortices 
 to result in quasi-two-dimensional (Q2D) flow,
 while a small rotation rate allows turbulence 
 to be three-dimensional (3D).
 It is found that
 the transition from the Q2D turbulent flow to the 3D turbulent flow
 and the reverse transition occur at different values of the rotation rates.
 At the intermediate rotation rates,
 bistability of these two statistically steady states is observed.
 Such hysteretic behavior is also observed 
for the variation of the amplitude of an external force.
\end{abstract}


\maketitle

Formation of columnar structures parallel to a rotation axis 
is one of the most fundamental
and distinctive phenomena in flows subject to rotation.
The emergence of columnar vortices in the rotating turbulence
makes a three-dimensional (3D) flow 
into a quasi-two-dimensional (Q2D) flow.
The Taylor--Proudman theorem has succeeded in explaining the cylindrical flow
in laboratory experiments and field observations in terms of the Taylor column.
However, the theorem cannot describe transitions between the Q2D 
and 3D flows,
because energy is exchanged between the Q2D mode and the 3D mode by nonlinear mechanisms~\cite{BELLET_GODEFERD_SCOTT_CAMBON_2006}.
The energy transfers to the Q2D modes were demonstrated by an instability analysis~\cite{doi:10.1063/1.858309,*:/content/aip/journal/pofa/5/3/10.1063/1.858651,*doi:10.1063/1.870022}
and weak-turbulence theory in the large-rotation limit~\cite{PhysRevE.68.015301}.
The Coriolis term breaks the parity invariance of the governing equation of the flow,
and introduces a scale-independent time scale
which induces two-dimensionalization at larger scales more effectively.
Therefore,
the Coriolis effect originates cyclone-anticyclone asymmetry
with enhanced stretching of cyclonic vorticity
and destabilization of anticyclonic one
due to the centrifugal instability and the vortex tilting~\cite{Bartello_Metais_Lesieur_1994,*Cambon_Benoit_Shao_Jacquin_1994,*:/content/aip/journal/pof2/20/8/10.1063/1.2966400}.

To classify the flow properties in rotating systems,
the Rossby number $\mathrm{Ro}$, 
which is the ratio between the linear and nonlinear time scales,
has been used~\cite{Hopfinger_Browand_Gagne_1982,*JFM273.1-29,*PhysFluids10.2895-2909}.
Note that though various definitions of $\mathrm{Ro}$ are used in literature,
the following facts are independent of its detailed definition.
When the Coriolis force is weak relative to turbulence,
i.e., $\mathrm{Ro}\gg1$,
the 3D Kolmogorov turbulence is obtained.
When $\mathrm{Ro}\sim 1$,
only cyclonic vortices appear at large scales,
and the flow becomes Q2D.
When $\mathrm{Ro}\ll1$,
both cyclonic and anticyclonic vortices appear,
and the flow fields are almost completely two-dimensionalized.
The transitions between the Q2D turbulence and the 3D turbulence 
by changing the system's rotation rate $\Omega$ 
were numerically studied~\cite{PhysRevLett.77.2467,*PhysRevE.90.023005}.
It was reported that
$\mathrm{Ro}$-dependence of turbulent statistics is not monotonic 
in the range $\mathrm{Ro}\sim1$,
where
the coherent vortices and inertial waves at small wave numbers
and the turbulence at large wave numbers
coexist~\cite{FLM:1318048,PhysRevE.91.043016}.
The two-dimensionalization and the cyclone-anticyclone asymmetry 
depend on the external forces and the boundary conditions (e.g., Ref.~\cite{AMR-14-1004}).

Recently, Ref.~\cite{Alexakis_2015} reported
a phase diagram for statistically steady states
of forced Taylor-Green flows in a rotating frame.
Four different steady states
in the parameter space spanned by the Reynolds numbers and the Rossby numbers
were numerically obtained by carrying out 184 simulations.
Subcritical behavior was implied by an abrupt transition
between the Q2D and 3D flows.
If it were a low-dimensional system,
one might expect a hysteresis in such transition.
Because the parameter space analysis was performed
with the same random initial condition for every parameter values in Ref.~\cite{Alexakis_2015},
the hysteretic behavior cannot be directly found.

It has been experimentally found that
there exists a parameter range 
where a high-torque state and a low-torque state are bistable 
and show a hysteretic behavior
in a highly turbulent Taylor--Couette flow~\cite{Huisman2014,*doi:10.1146/annurev-fluid-122414-034353,*PhysRevFluids.1.024401}.
A similar hysteretic behavior was observed also in rotating spherical Couette flow~\cite{doi:10.1063/1.3593465}.
Bistability and hysteresis between a stationary magnetic field and an oscillatory magnetic field in a low-dimensional phase space
were observed in a turbulent flow of liquid sodium~\cite{BERHANU_GALLET_MONCHAUX_BOURGOIN_ODIER_PINTON_PLIHON_VOLK_FAUVE_MORDANT}.
Note that 
the heteroclinic alternating transitions such as blocking in a rotating annulus~\cite{Weeks1598} associated with the Lorenz attractor,
where the state goes back and forth near the two {\em unstable\/} fixed points,
are different from the bistability.
Also note that the bistability does not necessarily indicate a hysteresis,
which requires a form of subcritical properties.
The dynamical systems theory developed in low-dimensional systems
has successfully been applied 
to the onset of turbulence, e.g., unstable periodic orbits in wall turbulence~\cite{KAWAHARA_KIDA_2001}.
The bifurcation structures embedded
in developed turbulence
would be a key to understand its nature.

Most of the hysteretic behavior in developed turbulence have been observed in the flows bounded by solid walls,
and the boundary condition plays an important role in the hysteretic behavior.
In this work,
a hysteretic behavior
in developed rotating turbulence
are numerically investigated in a periodic box.
As far as the authors are aware,
such behavior in the system not bounded by the solid walls has rarely been reported.
While the boundary condition and the forcing scheme in the present study
are the same as those in Ref.~\cite{Alexakis_2015},
the analytical methodologies for the parameter dependence of the flow patterns are different;
a flow field in the statistically steady state for a close parameter
is employed as an initial condition in this study,
whereas the random initial condition were used in Ref.~\cite{Alexakis_2015}.
The present approach is similar to the continuation of a solution
which enables us to track a branch in multistability
used in bifurcation analysis in low-dimensional dynamical systems.

We examine
the dependence of flow properties on $\Omega$
as well as the amplitude of the external force,
focusing on the transition between the Q2D flow and the 3D flow.
The micro-Rossby number defined below as well as the Taylor-scale Reynolds number
is the first candidate to characterize the flow field.
However, it is statistically defined by the flow itself,
and cannot be used as a control parameter.
The macro-Rossby number and the integral-scale Reynolds number can be control parameters,
but they do not well characterize the turbulent field.
As we will see below,
the transitions are hysteretic
owing to robustness of the large-scale columnar vortex
against the turbulent fluctuation.

The governing equations for the velocity $\bm{u}$ of the incompressible fluid
are the Navier--Stokes equation with the Coriolis term
and the divergence-free condition:
 \begin{align*}
  \frac{\partial \bm{u}}{\partial t}
  + (\bm{u} \! \cdot \! \nabla) \bm{u}
  + 2 \bm{\Omega} \times \bm{u}
  =
  - \nabla p
  + \nu \nabla^2 \bm{u}
  + \bm{f}
  ,
  \;
  \nabla \! \cdot \! \bm{u} = 0
 ,
\end{align*}%
where the centrifugal force is included in the pressure $p$.
The rotation vector $\bm{\Omega}=\Omega\bm{e}_z$ is assumed to be constant. 
The kinematic viscosity is expressed by $\nu$.
Note that the small wave-number drag is not added,
because it was reported that 
a statistically steady state can be achieved
even for the inversely cascading two-dimensional turbulence~\cite{PhysRevE.85.036315}.
The external force $\bm{f}$
is given by the three-dimensional two-component force of a steady TG type
$\bm{f}=f_0\allowbreak
(\cos k_{\mathrm{f}}x\sin k_{\mathrm{f}}y\sin k_{\mathrm{f}}z,
\allowbreak -\sin k_{\mathrm{f}}x\cos k_{\mathrm{f}}y \sin k_{\mathrm{f}}z,
\allowbreak 0)$,
where $k_\mathrm{f}=2$ is employed.
The TG flow has also been used as a model of many laboratory flows
(see Ref.~\cite{Alexakis_2015}).

In the present simulations,
the periodic boundary condition with the period $(2\pi)^3$ is employed.
The standard pseudo-spectral method with the aliasing removal
by the phase shift and the spherical truncation
is adopted for the nonlinear term,
and the numerical resolution is $512^3$.
The same value of $\nu$ is used for all the series of the simulations.
The Runge--Kutta--Gill method is adopted for the time integration,
while the linear terms are calculated analytically.
The characteristic length and time are selected
so that the period of the computational box is $2\pi$
and the Coriolis parameter $2\Omega$
and the amplitude of the external force $f_0$ are approximately $10$.

Let us first consider $\Omega$-dependence.
The rotation rate $\Omega$ is set
between $2.5$ and $7.5$ with $0.5$ increments or decrements,
while $f_0$ is fixed at $10$.
The rotation rate is increased or decreased by $0.5$
when the flow field for the previous rotation rate is in the statistically steady state.
The Zeman wave numbers $k_{\Omega}=(\Omega^3/\varepsilon)^{1/2}$,
where $\varepsilon$ is the energy dissipation rate,
are evaluated approximately as
$1.4$ for $\Omega=2.5$ and $16$ for $\Omega=7.5$.
The corresponding flows are 3D and Q2D.

\begin{figure}
  \begin{center}
    \includegraphics[scale=.95]{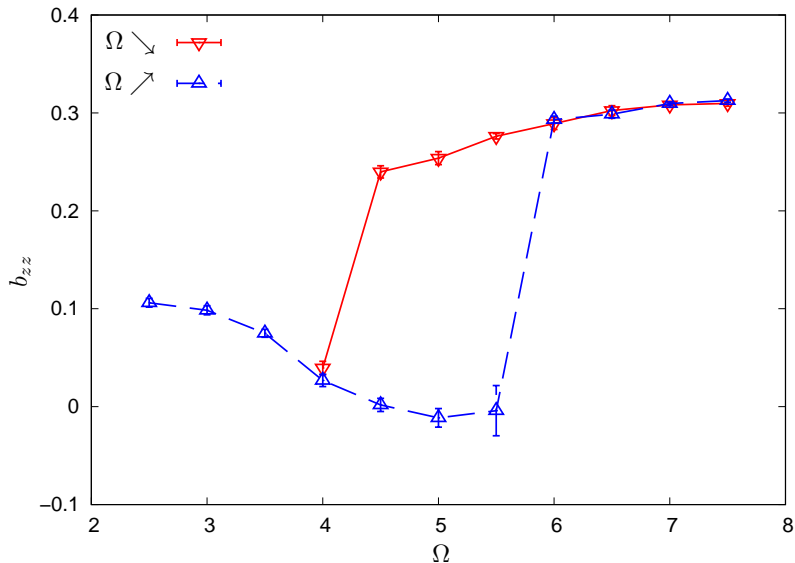}
  \end{center}
  \caption{
    Dependence of 
    $zz$ component of anisotropy tensor $b_{zz}$
    on $\Omega$.
    The solid curve ($\Omega\!\searrow$) and the dashed curve ($\Omega\!\nearrow$),
    respectively, show $b_{zz}$ during the decrease of $\Omega$
    and the increase of $\Omega$.
    The error bars represent the standard deviation due to the time variation.
  }
  \label{fig:hystrssbzzOmg}
\end{figure}

Since we are interested in the transitions between the Q2D turbulent flow and the 3D turbulent flow,
the dependence of the $zz$ component of the anisotropy tensor,
$b_{zz}=1/3-\langle u_z^2\rangle /\langle |\bm{u}|^2\rangle$,
on $\Omega$
is drawn in Fig.~\ref{fig:hystrssbzzOmg}
to evaluate the anisotropy of the flows.
These values are obtained in the statistically steady states.

At $\Omega=2.5$, $b_{zz}\approx0.1$,
where the non-zero value comes from the anisotropic external force of the TG type.
When the rotation rate is increased from $\Omega=2.5$ ($\Omega\!\nearrow$),
$b_{zz}$ decreases to almost $0$,
where the flow is 3D and almost isotropic.
It increases abruptly to about $0.3$
in the range of $5.5<\Omega<6$,
and the flow becomes Q2D and strongly anisotropic.
At $\Omega=7.5$, $b_{zz}\approx0.3$ owing to the strong rotation.
When the rotation rate is decreased from $\Omega=7.5$ ($\Omega\!\searrow$),
$b_{zz}$ slowly decreases.
In the range of $4<\Omega<4.5$,
$b_{zz}$ drops sharply, representing the abrupt transition
from the Q2D anisotropic flow to the 3D isotropic flow.
In the range of $4.5\lessapprox\Omega\lessapprox5.5$,
the two turbulent regimes
are bistable and show a hysteretic behavior.

\begin{figure}
\begin{center}
 \includegraphics[scale=.95]{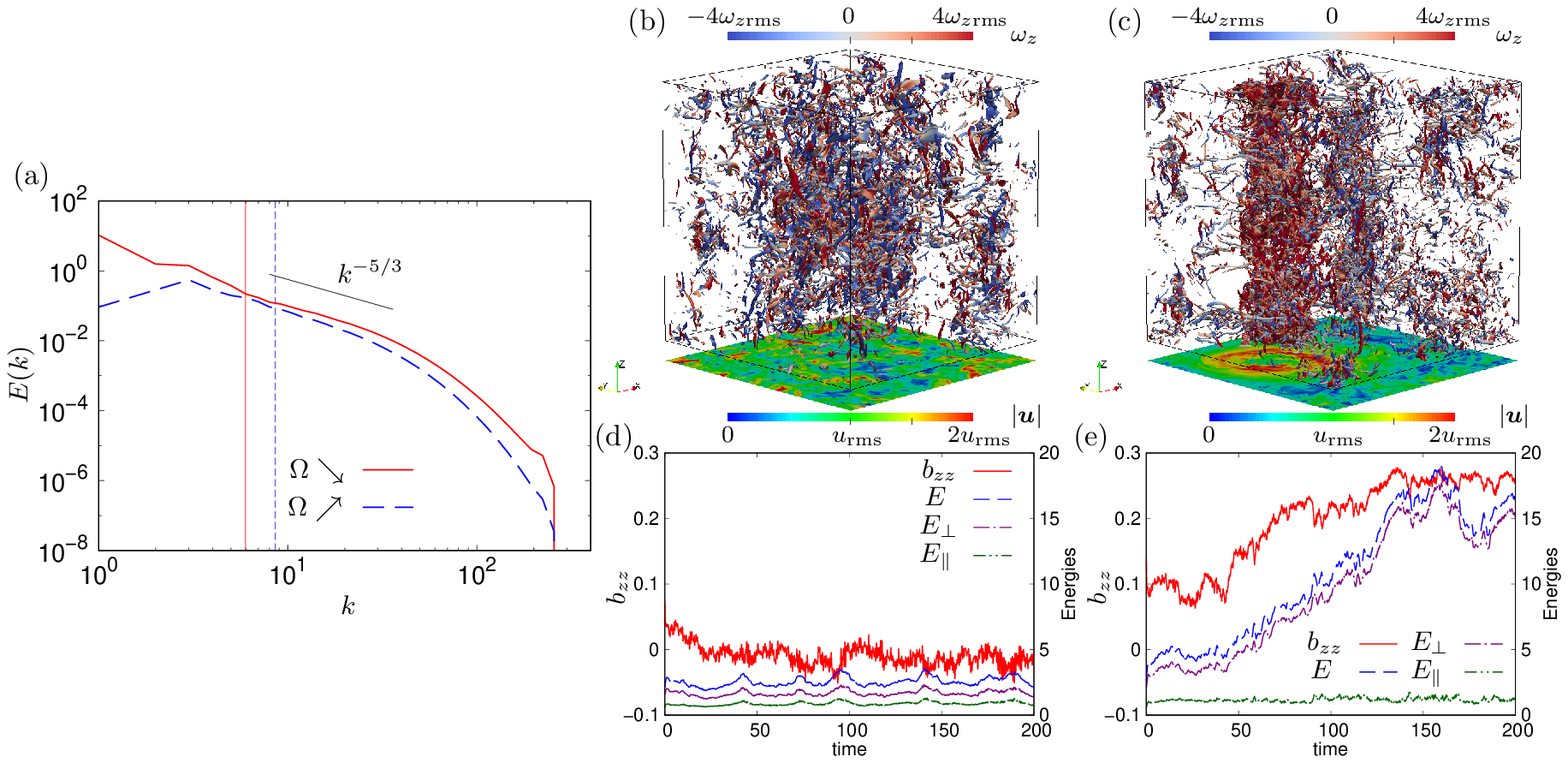}
\end{center}
 \caption{
 Bistability between the Q2D flow and the 3D flow at $\Omega=5$.
 Energy spectra (a). 
 The vertical lines represent the Zeman wave numbers.
 Isosurfaces of $|\bm{\omega}|=\pm3\sqrt{\langle|\bm{\omega}|^2\rangle}$
 colored by $\omega_z$
 and speed distribution on $z=0$ plane
 for lower branch $\Omega\!\nearrow$ (b) and for upper branch $\Omega\!\searrow$ (c).
 Time evolution of $b_{zz}$ (left axis)
 and total, perpendicular and parallel energies (right axis)
 for the interpolation weight $r=0.2$ (d) and for $r=0.3$ (e)
 between the Q2D flow and the 3D flow.
 }
 \label{fig:twostates5}
\end{figure}

To characterize the upper- and lower-branch flows
shown in Fig.~\ref{fig:hystrssbzzOmg},
the Fourier- and the real-space properties
at $\Omega=5$ are examined.
The one-dimensional energy spectra
are drawn in Fig.~\ref{fig:twostates5}(a).
For the lower-branch flow ($\Omega\!\nearrow$),
where $k_{\Omega}\approx8.6$,
the Kolmogorov spectrum $k^{-5/3}$ appears all over the inertial subrange.
For the upper-branch flow ($\Omega\!\searrow$),
while the Kolmogorov spectrum appears
in the range $k\gtrapprox k_{\Omega}\approx6.0$,
the energy spectrum at $k\lessapprox k_{\Omega}$ shows another turbulent state.
A significant difference between the upper- and lower-branch flows appears at the small wave numbers;
while the largest energy of the former appears 
at the scale of the external force, $\sqrt{3}k_\mathrm{f}\approx3.46$,
that of the latter appears at the largest scale $k=1$.

In the lower-branch flow,
the energy supplied by the external force is almost completely transferred
to the larger wave numbers as the 3D Kolmogorov turbulence.
In the upper-branch flow, on the other hand,
a part of the supplied energy is transferred to the smaller wave numbers
owing to the quasi-two-dimensionalization,
and the field reaches a statistically steady state 
when the accumulation at $k \approx 1$,
more precisely $(k_x, k_y, k_z)=(\pm1,0,0),(0,\pm1,0)$,
is built up.
The process of the accumulation is known as condensation.
The inverse cascade due to the quasi-two-dimensionalization
and the forward cascade from the accumulation to the large wave numbers
are considered to balance with each other
in the statistically steady state.
Similar discussions can be found in Ref.~\cite{Alexakis_2015}.

The Q2D flow has larger energy than the 3D flow all over the wave numbers.
Therefore, the two turbulent flows have different values of
the micro-Rossby number $\mathrm{Ro}_{\omega_z}=\omega_{z\mathrm{rms}}/(2\Omega)$
as well as the Taylor-scale Reynolds number $\mathrm{Re}_{\lambda}=(20E^2/(3\nu\varepsilon))^{1/2}$
under the identical parameters of the simulations:
$\mathrm{Ro}_{\omega_z}\approx2.5$ and $\mathrm{Re}_{\lambda}\approx490$ in the Q2D flow,
and $\mathrm{Ro}_{\omega_z}\approx1.8$ and $\mathrm{Re}_{\lambda}\approx110$ in the 3D flow.
Here, the subscript rms denotes the root-mean square, 
and $E$ is the total energy.
Note that the Taylor microscale for the isotropic turbulence
is used to obtain the values of $\mathrm{Re}_{\lambda}$,
because the Kolmogorov spectrum appears at the large wave numbers
in both Q2D and 3D flows.
The fact that
$\mathrm{Ro}_{\omega_z}$ in the Q2D flow is larger than that in the 3D flow
causes the non-monotonicity of the $\mathrm{Ro}$-dependence of turbulent statistics as reported in Ref.~\cite{FLM:1318048}.

Isosurfaces of vorticity norm $|\bm{\omega}|$ in the real space are drawn 
in Figs.~\ref{fig:twostates5}(b) and (c).
In the lower-branch flow, Fig.~\ref{fig:twostates5}(b),
no large-scale vortex is formed,
and we can observe only the small-scale 3D vortices.
On the other hand, 
in the upper-branch flow, Fig.~\ref{fig:twostates5}(c),
the isosurfaces show the cyclonic vortex aligned along the rotation axis,
which reminds us of the Taylor column.
The cyclonic vortex makes strongly sheared regions 
between itself and its images due to the periodic boundary condition.
In the strongly sheared regions,
cylindrical swarm composed of anticyclonic small-scale vortices is produced,
though they are weak.
In fact,
the cyclonic vortex makes a high-speed whirl 
as drawn on $z=0$ plane in Fig.~\ref{fig:twostates5}(c),
while the swarm does not.
The cyclonic vortex accompanied by inertial waves makes the upper-branch flow Q2D.

Even though the simulations were run long,
the possibility of transition between the Q2D flow and the 3D flow cannot be excluded.
To confirm the bistability at $\Omega=5$,
the existence of a basin of attraction between the Q2D flow and the 3D flow
is examined by starting from interpolated initial conditions.
The initial conditions are made
by the superposition of the Q2D flow and the 3D flow
$\bm{u}=r\bm{u}_{\mathrm{Q2D}}+(1-r)\bm{u}_{\mathrm{3D}}$,
where $r \in [0,1]$ is the weight for the interpolation between the Q2D flow and the 3D flow.
The simulations are performed for the weights $r=0.1,0.2,\cdots,0.9$
until each field reaches a statistically steady state.

The time evolutions of $b_{zz}$,
$E$,
perpendicular energy $E_{\perp}=\sum_{\bm{k}}|\widetilde{\bm{u}}_{\perp}|^2/2$
and parallel energy $E_{\|}=\sum_{\bm{k}}|\widetilde{u}_z|^2/2$
for $r=0.2$ and $0.3$ 
are drawn in Figs.~\ref{fig:twostates5}(d) and (e).
Here, $\widetilde{\bm{u}}_{\perp}$ is the Fourier coefficient
of the velocity component perpendicular to the rotation axis $\bm{u}_{\perp} = (u_x, u_y, 0)$.
In the simulation with $r=0.2$,
$b_{zz}$ decreases to around $0$,
while 
$b_{zz}$ increases to around $0.25$ in the simulation with $r=0.3$.
That is,
the boundary separating the Q2D flow and the 3D flow
exists in the range $0.2<r<0.3$,
and the Q2D flow and the 3D flow are bistable at $\Omega=5$.
The separatrix might not be a thin boundary
like the one which separates laminar and turbulent flows
at the onset of turbulence~\cite{doi:10.1143/JPSJ.70.703,*Schneider577}
but a thick and blurred region.

It is of interest to note that
both $E_{\|}$'s for $r=0.2$ and $0.3$ remain small
and the main difference appears in $E_{\perp}$.
Since $b_{zz}=1/3-E_{\|}/(E_{\perp}+E_{\|})$,
the transitions between the Q2D flow and the 3D flow
occur mainly in $\bm{u}_{\perp}$, i.e., $\omega_z$.
In other words, the fluctuation in the rotation direction, $u_z$,
is little affected by the rotation.

At the transition from the 3D flow to the Q2D flow, 
the coherent cyclonic vortex is formed
by overwhelming the external force
as well as turbulent fluctuation.
The external force in the present simulations
tries to develop the TG vortices
whose symmetries and scales are different from the coherent cyclonic vortex.
On the other hand,
at the reverse transition,
the coherent vortex collapses into pieces.
The formation and destruction of the large coherent vortex
cannot be continuous for the variation of $\Omega$.
Therefore,
the transitions exhibit the hysteresis.

\begin{figure}
  \begin{center}
    \includegraphics[scale=.95]{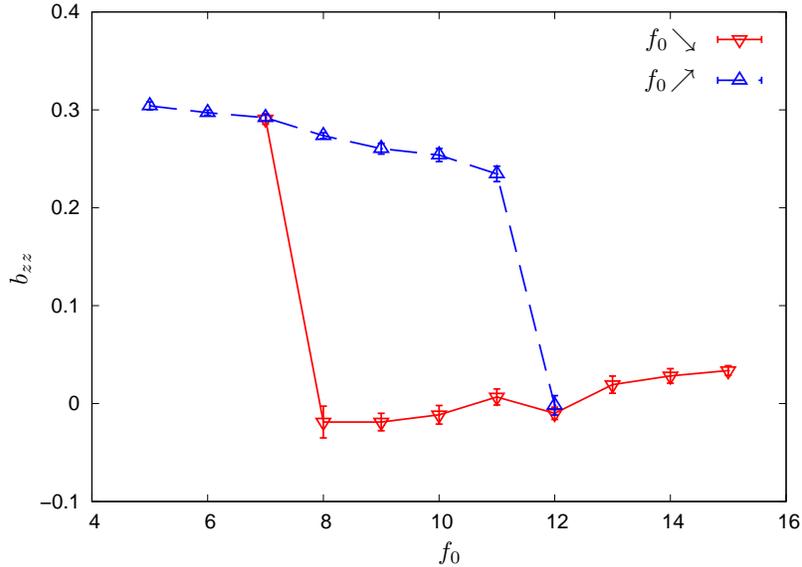}
  \end{center}
  \caption{
    $f_0$-dependence of $b_{zz}$.
 See also the caption of Fig.~\ref{fig:hystrssbzzOmg}.
 }
  \label{fig:hystrssbzzf0}
\end{figure}

The dependence of $b_{zz}$
on the amplitude of the external force $f_0$,
which is increased with $1$ increments or decreased with $1$ decrements,
is drawn in Fig.~\ref{fig:hystrssbzzf0}.
Following the results of the $\Omega$-dependence
in Fig.~\ref{fig:hystrssbzzOmg},
we here investigate the cases with $\Omega=5$ as representative.
Similar to Fig.~\ref{fig:hystrssbzzOmg},
the Q2D flow and the 3D flow are, respectively, observed at small $f_0$ and large $f_0$,
and the two turbulent flows are bistable and hysteretic at intermediate $f_0$.
Obviously, this two-dimensionalization is due to the cyclonic vortex along the rotation axis.

\begin{figure}
  \begin{center}
    \includegraphics[scale=.9]{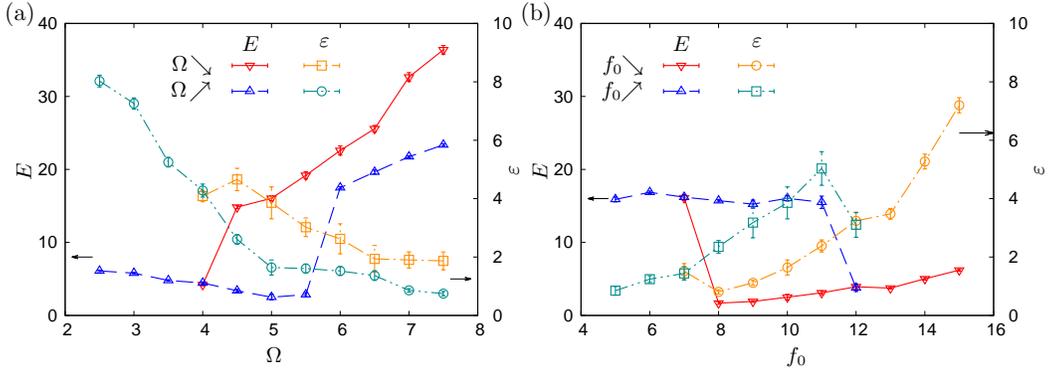}
  \end{center}
  \caption{
 $\Omega$-dependence (a)
 and $f_0$-dependence (b)
 of total energy (left axis) and energy dissipation rate (right axis).
  }
  \label{fig:hysteresisED}
\end{figure}

To analyze the properties of the above hysteretic behaviors
at the large and small scales,
the dependences of the total energy $E$
and the energy dissipation rate $\varepsilon$
on $\Omega$ and $f_0$
are shown in Fig.~\ref{fig:hysteresisED}.
These figures show different behaviors
between the $\Omega$- and $f_0$-dependences even qualitatively.
The transitions of $E$ and $\varepsilon$ show the bistability
in $4.5\lessapprox\Omega\lessapprox5.5$ and $8\lessapprox f_0\lessapprox11$
as observed in Figs.~\ref{fig:hystrssbzzOmg} and \ref{fig:hystrssbzzf0}.
However, both $E$ and $\varepsilon$ have different values at large $\Omega$
where the flow is Q2D.
No clear jump in $5.5\leq\Omega\leq6$ can be seen
in $\varepsilon$ for increasing $\Omega$ 
in Fig.~\ref{fig:hysteresisED}(a).
In other words, small-scale dynamics is insensitive to $\Omega$,
which is consistent with the fluctuation in the rotation direction,
$E_{\|}$,
observed in Figs.~\ref{fig:twostates5}(d) and (e).

The fact that
the discrepancy between the two branches at large $\Omega$ is larger
than the turbulent fluctuation
and the hysteresis loop is not closed
implies the multiplicity of the statistically steady states of the Q2D flows.
The multiplicity is confirmed by finding the separatrix of the superposition of the two flows
in the same way to demonstrate the bistability between the Q2D flow and the 3D flow at $\Omega=5$ (Figs.~\ref{fig:twostates5}(d) and (e)).
It should be noted that
$b_{zz}$ shown in Fig.~\ref{fig:hystrssbzzOmg}, 
which is a non-dimensionalized quantity
composed of the ratio of the parallel energy to the total one,
is insensitive to the variation of the energies themselves.

It may appear to be counterintuitive
that the total energy for the small external force
is larger than that for the large external force
as shown in Fig.~\ref{fig:hysteresisED}(b).
It can be explained by the formation process of the large-scale cyclonic vortex.
When the external force is weak,
the Coriolis force is strong relative to turbulence intensity.
Then,
the large-scale columnar vortex is formed,
and it makes the large energy accumulation near $k_z = 0$.
Conversely,
when the external force is strong,
the Coriolis force is relatively weak.
Then, 
the turbulent flow is 3D,
and all the energy supplied by the external force is cascaded forwardly.
As a result, the energy does not accumulate near $k_z=0$,
 and the total energy is small.

\begin{figure}
 \begin{center}
  \includegraphics[scale=.9]{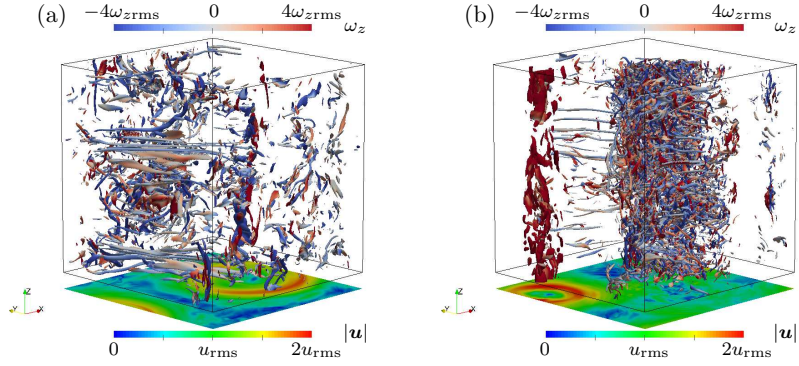}
 \end{center}
 \caption{
 Isosurface of $|\bm{\omega}|=\pm3\sqrt{\langle|\bm{\omega}|^2\rangle}$
 colored by $\omega_z$
 and speed distribution on $z=0$ plane
 at $\Omega=7.5$.
 (a): lower branch, 
 and
 (b): upper branch.
 }
  \label{fig:isosrfcO75}
\end{figure}

Lastly, to investigate the characteristics of the multiplicity
at the large $\Omega$, 
the isosurfaces of $|\bm{\omega}|$ for each branch at $\Omega=7.5$
are shown in Fig.~\ref{fig:isosrfcO75}(a).
Obviously, both flows have the Q2D structures as indicated
in Fig.~\ref{fig:hystrssbzzOmg}.
While the upper-branch flow
has the strong slender cyclonic vortex accompanied
by the swarm of anticyclonic small vortices,
the cyclonic vortex in the lower-branch flow is weak and fat,
and there is no room for the anticyclonic swarm to grow.
These flow patterns as well as the total energy and the energy dissipation rate
reveal the distinctive features
degenerated in the representation in terms of $b_{zz}$.
The Q2D flow might have more multiplicities other than those shown here.

Comparing with the phase diagram in Ref.~\cite{Alexakis_2015},
we recognize that most of the parameter values in the present study
fall into the quasi-2D condensates,
while some are located narrowly in the range of intermittent bursts and weakly rotating flows.
The quasi-2D condensates and the weakly rotating flows in Ref.~\cite{Alexakis_2015} respectively correspond to the Q2D flow and the 3D flow reported here.
An abrupt change in energy due to variation of Rossby number
between the Q2D flow and the 3D flow was observed,
and the subcritical behavior was implied by the abrupt change.
It should, however, be noted that
the initial conditions used in Ref.~\cite{Alexakis_2015}
were not the solutions for other parameters.
In this Rapid Communication,
the Q2D and 3D branches were traced by continuing the solution
to find whether the subcritical behavior results in
the hysteretic behavior or the heteroclinic alternating transitions
in large turbulent fluctuations.
The intermittent bursts are not observed in this study,
but intermittent growths of the total energy
which are caused by the energy transfers to the small wave numbers for a short time
appear in the 3D flow as recognized in Fig.~\ref{fig:twostates5}(d).
The intermittent growths in the 3D flow
are more frequent and not so strong
than the intermittent bursts observed in Ref.~\cite{Alexakis_2015}.

In summary, 
the $\Omega$-dependence and the $f_0$-dependence of turbulent flows
were investigated by numerically simulating the Navier--Stokes equation
with the Coriolis term under the steady forcing of the TG type.
The hysteretic behavior
between the Q2D flows observed at large $\Omega$'s and the 3D flows at small $\Omega$'s
was found.
This hysteretic behavior stems from 
the robustness of the large-scale cyclonic columnar vortex.
The hysteretic behavior between the Q2D flow and the 3D flow exists
for a finite bounded area in $(\Omega,f_0)$.
Although the flow properties had been classified
simply by using the micro-Rossby number~\cite{FLM:1318048}
and summarized in a phase diagram in Ref.~\cite{Alexakis_2015},
the present results demonstrate
that the selection of the flow structures
depends also on the initial conditions.
The hysteresis brings the complexity of $\mathrm{Ro}$-dependence
at $\mathrm{Ro}\sim1$.

This hysteretic behavior robustly exists
against the large fluctuation of the fully developed turbulence
whose energy spectra show the $-5/3$ power law. 
We also performed preliminary simulations
in which flows are excited by a white random force~\cite{supplementalmaterial}. 
The hysteresis and the bistability between the Q2D turbulent flow and the 3D turbulent flow
are observed also for such antithetical forcing,
though the range of the bistable parameters is much smaller.
The existence of the universal mechanism 
for the emergence of multiple flow patterns
in turbulence
is expected.
Dependence of the hysteretic behavior on the forcing types,
the multiplicity of the Q2D flow,
and the formation mechanism of the hysteretic behavior
will be reported elsewhere.

\begin{acknowledgments}
 Numerical computation in this work was carried out
 at the Yukawa Institute Computer Facility, Kyoto University
 and Research Institute for Information Technology, Kyushu University.
 This work was partially supported by JSPS KAKENHI Grant
 No.~15K17971 and No.~16K05490.
\end{acknowledgments}

%
\end{document}